# Tuning Fermi Levels in Intrinsic Antiferromagnetic Topological Insulators $MnBi_2Te_4$ and $MnBi_4Te_7$ by Defect Engineering and Chemical Doping

Mao-Hua Du[1,*], Jiaqiang Yan[1], Valentino R. Cooper[1], and Markus Eisenbach[2]

[1]Materials Science and Technology Division and [2]Center for Computational Sciences,

Oak Ridge National Laboratory, Oak Ridge, TN 37831, USA

*mhdu@ornl.gov

## Abstract

$MnBi_2Te_4$ and $MnBi_4Te_7$ are intrinsic antiferromagnetic topological insulators, offering a promising materials platform for realizing exotic topological quantum states. However, high densities of intrinsic defects in these materials not only cause bulk metallic conductivity, preventing the measurement of quantum transport in surface states, but may also affect magnetism and topological properties. In this paper, we show by density functional theory calculations that the strain induced by the internal heterostructure promotes the formation of large-size-mismatched antisite defect $Bi_{Mn}$ in $MnBi_2Te_4$; such strain is further enhanced in $MnBi_4Te_7$, giving rise to even higher $Bi_{Mn}$ density. The abundance of intrinsic $Bi_{Mn}$ donors results in degenerate n-type conductivity under the Te-poor growth condition. Our calculations suggest that growths in a Te-rich condition can lower the Fermi level, which is supported by our transport measurements. We further show that the internal strain can also enable efficient doping by large-size-mismatched substitutional $Na_{Mn}$ acceptors, which can compensate $Bi_{Mn}$ donors and lower the Fermi level. Na doping may pin the Fermi level inside the bulk band gap even at the Te-poor limit in $MnBi_2Te_4$. Furthermore, facile defect formation in $MnSb_2Te_4$ and its implication in Sb doping in $MnBi_2Te_4$ as well as the defect segregation in $MnBi_4Te_7$ are discussed. The defect engineering and doping strategies proposed in this paper will stimulate further studies for improving synthesis and for manipulating magnetic and topological properties in $MnBi_2Te_4$, $MnBi_4Te_7$, and related compounds.

# I. Introduction

Topological insulators (TI) have bulk gaps but metallic surface/edge states with linear dispersion protected by the time reversal symmetry (TRS)[1-4]. The presence of a long-range magnetic order in a TI can break the TRS and opens an exchange gap in the surface states; this could lead to a variety of exotic topological quantum states, including quantum anomalous Hall (QAH) effect and axion insulator states[1,2]. Extensive research has been carried out to search for materials platforms that enable the observation of these quantum phenomena[5-8].

$MnBi_2Te_4$ and related van der Waals (vdW) heterostructures $(MnBi_2Te_4)(Bi_2Te_3)_n$ have recently emerged as a new class of intrinsic antiferromagnetic (AFM) TIs.[9-18] $MnBi_2Te_4$ consists of stacked septuple layers (SLs) of Te-Bi-Te-Mn-Te-Bi-Te as shown in Figure 1(a). $Bi_2Te_3$ quintuple layers (QLs) can be inserted between $MnBi_2Te_4$ SLs to form $(MnBi_2Te_4)(Bi_2Te_3)_n$ [e.g., $MnBi_4Te_7$ (n = 1) and $MnBi_6Te_{10}$ (n = 2)] [see Figure 1(b) for the crystal structure of $MnBi_4Te_7$].[19] In these compounds, Mn ions have the intralayer FM coupling while the interlayer coupling is AFM, forming A-type AFM ordering.[20,21] The Dirac surface states have been demonstrated in $MnBi_2Te_4$ and $MnBi_4Te_7$ by angle-resolved photoemission spectroscopy (ARPES) and theoretical calculations.[9-15,18] The QAH effect[16] and the axion insulator state[17] have recently been reported in $MnBi_2Te_4$ with odd and even number of SLs, respectively, but have not yet been demonstrated in $MnBi_4Te_7$.

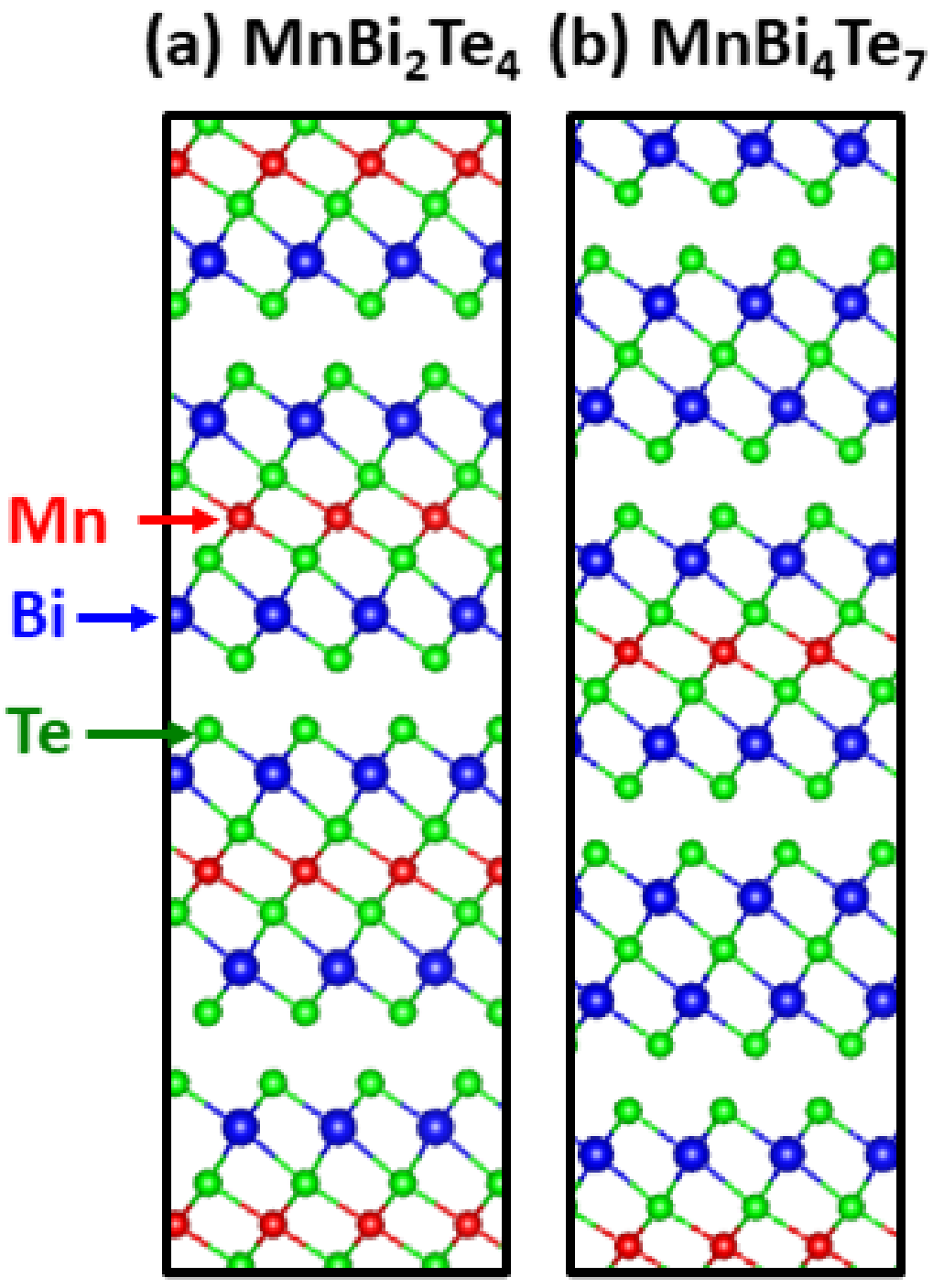

Figure 1. Crystal structures of (a) $MnBi_2Te_4$ and (b) $MnBi_4Te_7$.

The key to the experimental observation of the QAH effect and the axion state is the control of the Fermi level, which needs to be within the bulk band gap. This is, nevertheless, challenging and not realized in $MnBi_2Te_4$ and $MnBi_4Te_7$ samples, in which the Fermi level is above the conduction band minimum (CBM) according to the ARPES. Transport measurements show the behavior of degenerate n-type semiconductors,[9,10,12,13,20,22,23] consistent with ARPES results. A frequently used approach to tune the Fermi level is applying a gate voltage to a thin film or a flake in a FET device configuration.[16,17] However, a FET setup is not feasible in all experimental measurements and the Fermi level could be far from the band gap, requiring a very large gate voltage[16], not practical for many device applications. Therefore, the availability of insulating single crystals suitable for various experimental measurement and device environments is highly desirable. Doping of $MnBi_2Te_4$ by replacing Bi with isovalent Sb has also been shown to lower the Fermi level, leading to a transition from n- to p-type conductivity in Sb doped $MnBi_2Te_4$.[24-26] The Fermi level is tuned to near the valence band maximum (VBM) after replacing about 30% Bi by Sb.[25,26] The drawback of Sb doping is the reduction of the SOC effect and the band gap,[25] which positions the Dirac point closer to band edges, making the transport measurement of surface states more difficult. The AFM $MnSb_2Te_4$ was shown to be topologically trivial.[25] The Sb doping may also introduce more antisite defects as we discuss in this paper. Doping of $MnBi_2Te_4$ by electrically active acceptor dopants has not been reported. Previous studies on $Bi_2Se_3$ show that Ca doping can reduce n-type carrier density and achieve n-to-p type transition in $Bi_2Se_3$.[27,28]

$MnBi_2Te_4$, $MnBi_4Te_7$ and $MnBi_6Te_{10}$ have high density of defects especially the latter two. Among these three compounds, $MnBi_2Te_4$ seems to be closest to being stoichiometric with a few percent of antisite disorder,[20,29] although there is a report showing more than 15% Mn deficiency.[21] The Mn deficiency may be sample-dependent. $MnBi_4Te_7$ and $MnBi_6Te_{10}$ are structurally related to $MnBi_2Te_4$ with the same $MnBi_2Te_4$ SLs, but have severe Mn deficiencies, which appear to increase with the addition of more $Bi_2Te_3$ QLs in the unit cell. The Mn deficiencies were estimated to be 15%-20% in $MnBi_4Te_7$ and 19%-25% in $MnBi_6Te_{10}$.[14,21,30] Such high defect densities not only affect the Fermi level but could also influence the magnetic and topological properties. For example, a high density of magnetic defects may affect the surface magnetism and the opening of the surface energy gap. Furthermore, the two possible surface terminations by either magnetic $MnBi_2Te_4$ SLs or nonmagnetic $Bi_2Te_3$ QLs in $MnBi_4Te_7$ and $MnBi_6Te_{10}$ can give rise to two different sets of topological surface states.[14,15,31,32] It is unclear whether the defect concentration varies between SLs and QLs. A detailed study of the defect distribution in $MnBi_4Te_7$ is highly desirable for understanding and tuning topological surface states.

Despite the critical importance of defect management for synthesis and for the observation of topological quantum states, the defects that cause the nonstoichiometry and the underlying mechanisms behind the apparent different defect chemistries in $MnBi_2Te_4$, $MnBi_4Te_7$ and $MnBi_6Te_{10}$ are not well understood. In this paper, we study defect and dopant properties in $MnBi_2Te_4$ and $MnBi_4Te_7$ under different growth conditions by density functional theory (DFT) calculations. Defects in $MnSb_2Te_4$ are also studied as heavy doping of $MnBi_2Te_4$ by Sb can effectively tune the Fermi level. We show that the antisite defect $Bi_{Mn}$ is the dominant intrinsic donor defects in both $MnBi_2Te_4$ and $MnBi_4Te_7$. $Bi_{Mn}$ cannot be completely compensated by any intrinsic acceptors at the Te-poor limit, leading to the behavior of degenerate n-type semiconductors for both compounds. Adopting a Te-rich condition is predicted to lower the Fermi level. These results are consistent with our transport measurements on $MnBi_2Te_4$ samples grown with different excess Te contents. In contrast, our defect calculations show that $MnSb_2Te_4$ is intrinsically p-type with $Mn_{Sb}$ as the dominant acceptor defect. We predict increased defect concentrations from $MnBi_2Te_4$, $MnBi_4Te_7$, to $MnSb_2Te_4$ based on our calculations. As a result, heavy Sb doping of $MnBi_2Te_4$ should induce a higher density of defects, and a high density of magnetic defects $Mn_{Sb}$ in $MnSb_2Te_4$ can affect magnetic ordering in $MnSb_2Te_4$. Several alkali-metal (Li, Na, K) and alkali-earth-metal (Be, Mg, Ca) dopants are studied in both $MnBi_2Te_4$ and $MnBi_4Te_7$, and Na is found to be the most effective p-type dopant, which can compensate $Bi_{Mn}$ donors, pinning the Fermi level within the band gap even at the Te-poor limit in $MnBi_2Te_4$. We also find that the important magnetic defect $Mn_{Bi}$ in $MnBi_4Te_7$ prefers to form in $Bi_2Te_3$ QLs and that the $MnBi_2Te_4$ SL termination has a lower surface energy than the $Bi_2Te_3$ QL termination. The surprisingly low formation energies of large-size-mismatched antisite defects and dopants and the higher defect concentrations in $MnBi_4Te_7$ than in $MnBi_2Te_4$ are explained in the context of strain induced by intercalating MnTe within the $Bi_2Te_3$ layer.

## II. Results

### A. Enthalpy of formation and chemical Potentials

The calculated enthalpies of formation for the reactions MnTe + $Bi_2Te_3$ → $MnBi_2Te_4$, MnTe + $2Bi_2Te_3$ → $MnBi_4Te_7$, $MnBi_2Te_4$ + $Bi_2Te_3$ → $MnBi_4Te_7$, and MnTe + $Sb_2Te_3$ → $MnSb_2Te_4$ are -5 meV, 5 meV, 10 meV, and 6 meV, respectively. (A negative enthalpy formation indicates exothermic reaction.) These enthalpies of formation are all close to zero (within ± 10 meV), suggesting that synthesis of these compounds can be challenging and that the entropy contribution at elevated temperatures is important to the growth of these ternary compounds. In this paper, the above enthalpies of formation are approximated to zero. As a result, the calculated chemical potential range under thermal equilibrium is represented approximately by a line segment (between points A and B in Figure 2) rather than a typical polygon because the phase boundaries between the ternary phase (e.g., $MnBi_2Te_4$) and two binary secondary phases (e.g., MnTe and $Bi_2Te_3$) overlap when the enthalpy of formation is approximated to zero. The details involved in the calculation of Figure 2 is given in Section V.

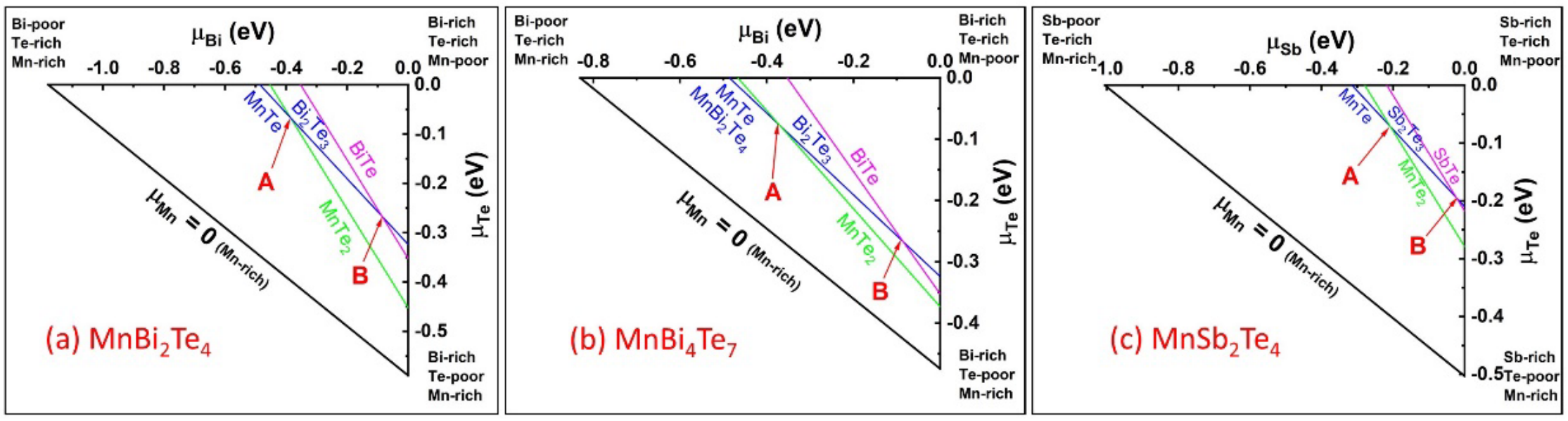

Figure 2. Calculated chemical potential ranges of constituent elements in (a) $MnBi_2Te_4$, (b) $MnBi_4Te_7$, and (c) $MnSb_2Te_4$, respectively. Points A and B correspond to the Te-rich/cation-poor and the Te-poor/cation-rich limits.

### B. Intrinsic defects in $MnBi_2Te_4$ and $MnBi_4Te_7$

Formation energies of intrinsic point defects, including vacancies, interstitials, and antisites, in $MnBi_2Te_4$ and $MnBi_4Te_7$ (with the AFM ordering) are shown in Figures S1 and S2, respectively; the most important low-energy defects are shown in Figures 3 and 4. Some defects can form on several inequivalent sites (such as $Mn_{Bi}^-$ and $Bi_{Te}^-$ in $MnBi_4Te_7$); formation energies for these defects are shown for the energetically most favorable site. As seen in Figures 3 and 4, the antisite defect $Bi_{Mn}^+$ is the most stable donor defect in both $MnBi_2Te_4$ and $MnBi_4Te_7$. The most important acceptor defects are also antisite defects, i.e., $Mn_{Bi}^-$ and $Bi_{Te}^-$. $Mn_{Bi}^-$ is the most stable acceptor defect at the Te-rich limit in both compounds. At the Te-poor limit, both $Mn_{Bi}^-$ and $Bi_{Te}^-$ have low formation energies; the former is slightly more stable in $MnBi_4Te_7$ while the latter is slightly more stable in $MnBi_2Te_4$. $Bi_{Te}^{-}$ has previously been identified as the dominant acceptor defect at the Te-poor limit in $Bi_2Te_3$,[3,33,34] consistent with our results in $MnBi_2Te_4$ and $MnBi_4Te_7$. Vacancies and interstitials all have high formation energies, as shown in Figures S1 and S2, and are not important defects in $MnBi_2Te_4$ and $MnBi_4Te_7$.

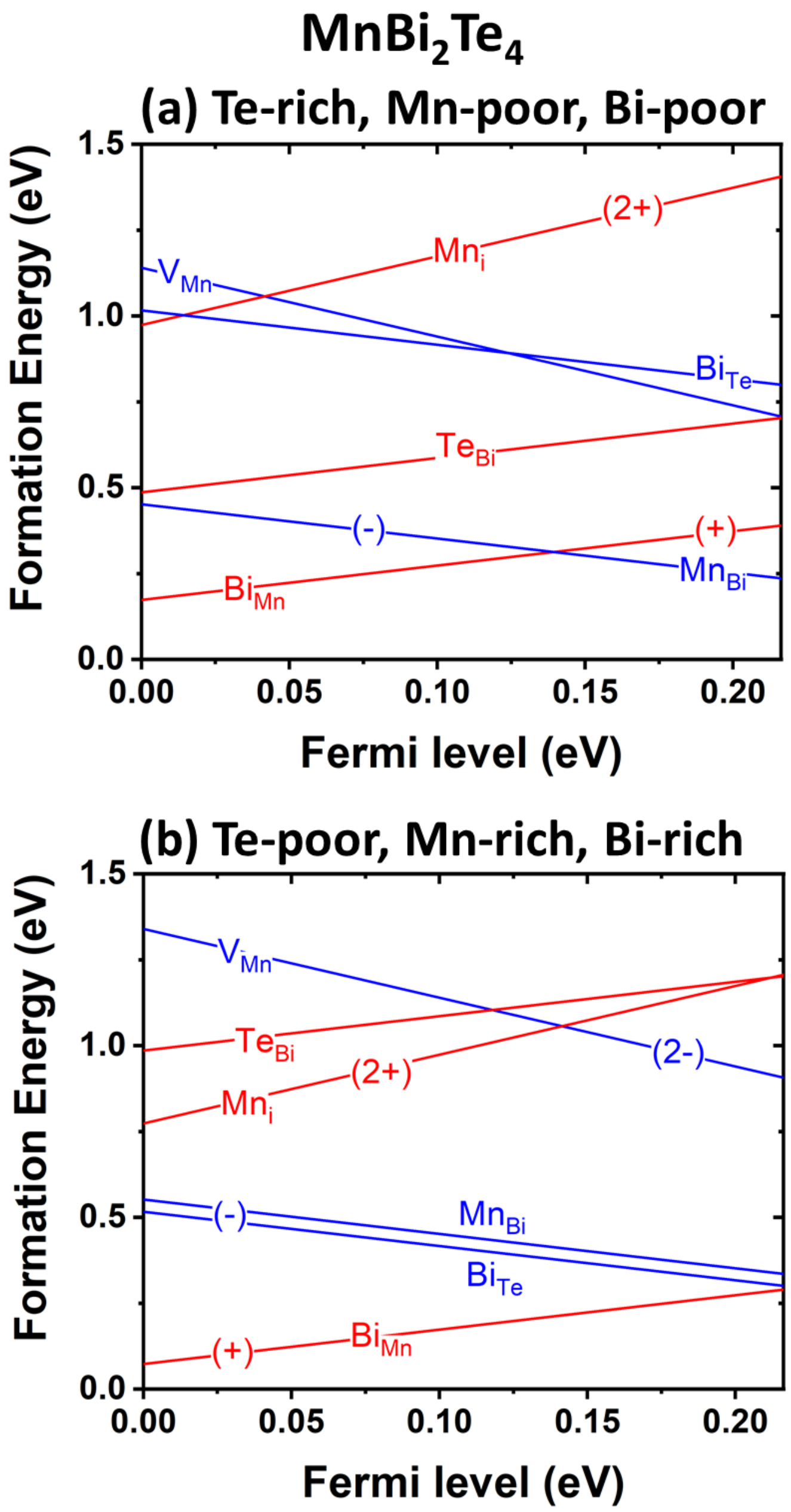

Figure 3. Calculated formation energies of intrinsic defects as functions of the Fermi level (varied from the VBM to the CBM) at the Te-rich/cation-poor (a) and Te-

poor/cation-rich (b) limits [corresponding to points A and B in Figure 2(a), respectively] in $MnBi_2Te_4$. The slope of a formation energy line indicates the charge state of the defect as selectively shown.

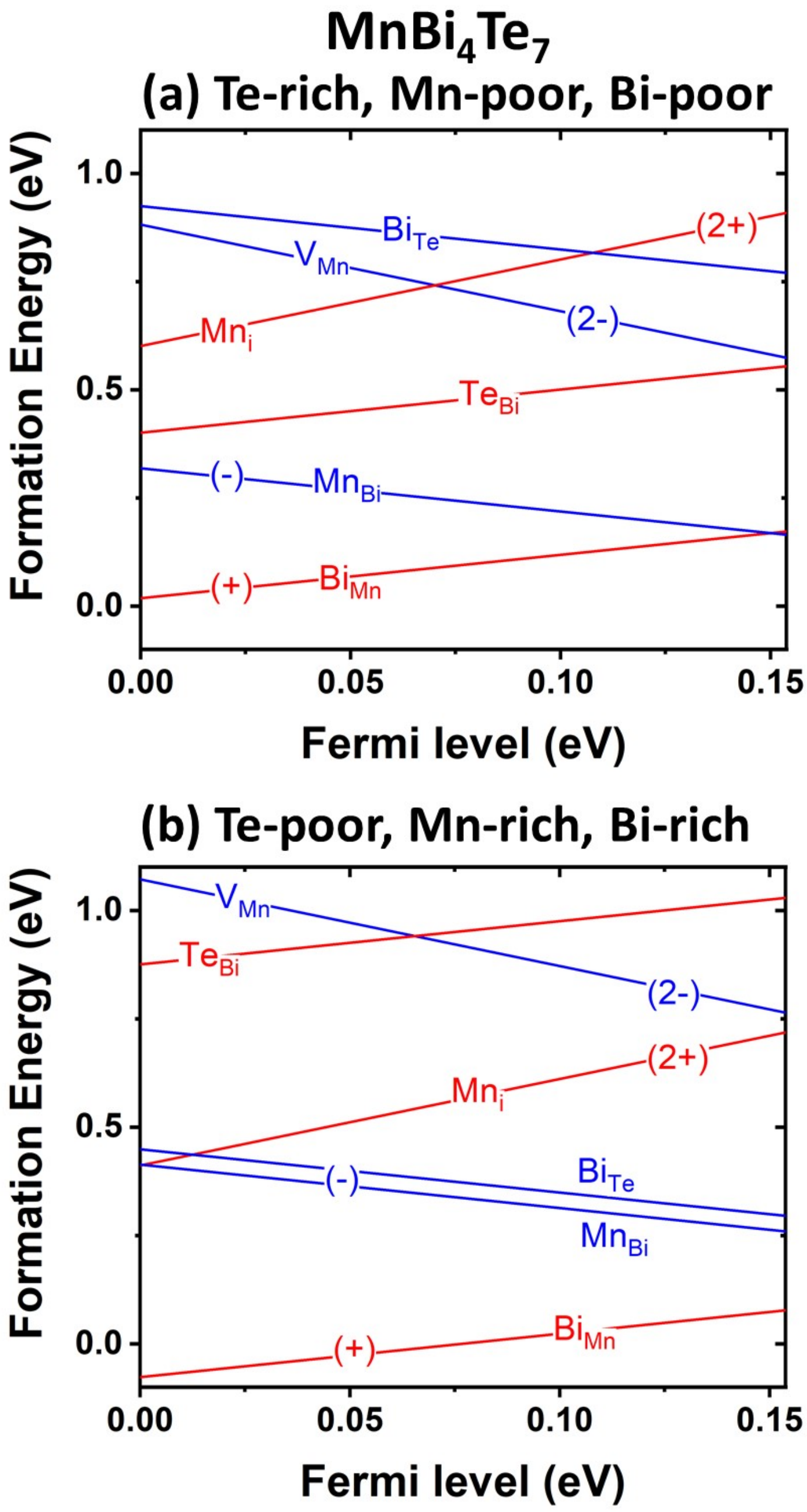

Figure 4. Calculated formation energies of intrinsic defects as functions of the Fermi level (varied from the VBM to the CBM) at the Te-rich/cation-poor (a) and Te-poor/cation-rich (b) limits [corresponding to points A and B in Figure 2(b), respectively] in $MnBi_4Te_7$. The slope of a formation energy line indicates the charge state of the defect as selectively shown.

### C. Fermi level and defect concentrations in $MnBi_2Te_4$ and $MnBi_4Te_7$

Based on calculated defect formation energies, we further estimate defect and free carrier densities as well as the Fermi level assuming thermal equilibrium and charge neutrality. We calculated defect densities at T = 585 °C (the melting temperature of $Bi_2Te_3$) by solving Eqs. 9-12 self-consistently. Since the magnetic and topological properties are measured at low temperatures, the Fermi level and the free carrier density at 2 K were then calculated by solving Eqs 9-12 again with the defect densities fixed at the values calculated at 585 °C. This simulation assumes that defects created at the growth temperature are frozen in the lattice during the subsequent cooling. The implication of this assumption is further discussed below. If there are multiple inequivalent lattice sites for the formation of a defect, formation energies and populations of the defect on all possible sites are considered.

Defect formation energies are strongly affected by elemental chemical potentials, which vary under different experimental growth conditions. Both $MnBi_2Te_4$ and $MnBi_4Te_7$ are n-type at the Te-poor limit and p-type at the Te-rich limit as shown in Table I. The significant amount of $Bi_{Te}$ observed on the $MnBi_2Te_4$ surface by scanning transmission microscopy (STM) [20,29] suggests that the growth condition is likely Te-poor because $Bi_{Te}$ is abundant under only the Te-poor conditions as shown by the calculated formation energies in $MnBi_2Te_4$ (Figure 3) and $MnBi_4Te_7$ (Figure 4). Table I also shows that defect densities and the Mn deficiency increases from $MnBi_2Te_4$ to

$MnBi_4Te_7$, consistent with the trend observed experimentally. [14,21,30]

Table I. Calculated Fermi level ($\varepsilon_f$), free carrier density (and the type of the carrier), as well as densities of the most important intrinsic defects (and concentrations in atomic percent) at both Te-rich and Te-poor limits in $MnBi_2Te_4$ and $MnBi_4Te_7$. The Mn:Bi:Te composition ratios calculated based on defect densities are also shown. $\mathrm{Bi_{Mn}^{+}}$ and $\mathrm{Mn_{Bi}^{-}}$ are important at both Te-rich and Te-poor limits while $\mathrm{Bi_{Te}^{-}}$ is important at only the Te-poor limit. The Fermi level is referenced to the valence band maximum ($E_V$) or the conduction band minimum ($E_C$). The calculated band gaps of $MnBi_2Te_4$ and $MnBi_4Te_7$ are 0.22 eV and 0.15 eV, respectively. Defect densities are calculated at 858.15 K (585 °C) while free carrier densities are calculated at 2 K. The unit of all densities is $cm^{-3}$.

| | $MnBi_2Te_4$ | | $MnBi_4Te_7$ | |
|---|---|---|---|---|
| | Te-rich | Te-poor | Te-rich | Te-poor |
| $\varepsilon_f$ | $E_V$ - 0.01 eV | $E_C$ + 0.03 eV | $E_V$ + 0.06 eV | $E_C$ + 0.09 eV |
| Carrier density | $2.3\times10^{19}$ (p) | $8.6\times10^{18}$ (n) | $9.6\times10^{17}$ (p) | $6.2\times10^{19}$ (n) |
| $[\mathrm{Bi_{Mn}^{+}}]$ | $8.1\times10^{19}$ (1.8%) | $1.5\times10^{20}$ (3.4%) | $4.0\times10^{20}$ (15.7%) | $6.0\times10^{20}$ (23.9%) |
| $[\mathrm{Mn_{Bi}^{-}}]$ | $1.1\times10^{20}$ (1.2%) | $5.6\times10^{19}$ (0.6%) | $4.1\times10^{20}$ (4.0%) | $2.7\times10^{20}$ (2.6%) |
| $[\mathrm{Bi_{Te}^{-}}]$ | | $9.0\times10^{19}$ (0.5%) | | $2.7\times10^{20}$ (1.5%) |
| Mn:Bi:Te | 1.01 : 1.99 : 4 | 0.98 : 2.05 : 4 | 1.00 : 4.00 : 7 | 0.88 : 4.31 : 7 |

At the Te-poor limit, the calculated Fermi levels in $MnBi_2Te_4$ ($E_C$ + 0.03 eV) and $MnBi_4Te_7$ ($E_C$ + 0.09 eV) are both above the CBM, indicating degenerate n-type conductivities, consistent with ARPES and transport measurements.

[9,10,12,13,20,22,23] Adopting a Te-rich condition lowers the Fermi level in both compounds as shown in Table I. To verify this, we grew $MnBi_2Te_4$ single crystals out of $Bi_2Te_3$ flux but with extra Te added. As shown in Figure S3, the electron density decreases with increasing extra Te content in the starting materials, in agreement with the trend found in our calculations. More detailed characterization of these crystals is in progress and will be reported elsewhere.

Despite the agreement in the Fermi level trend between theory and experiment, the Fermi level measured by ARPES is higher (~ $E_C$ + 0.2 eV in both $MnBi_2Te_4$ and $MnBi_4Te_7$) [12,31] than the calculated values. The highest calculated Fermi level is obtained at the Te-poor limit ($E_C$ + 0.03 eV in $MnBi_2Te_4$ and $E_C$ + 0.09 eV in $MnBi_4Te_7$). The integration of the DOS from the CBM to $E_C$ + 0.2 eV in $MnBi_2Te_4$ yields a free electron density of $1.6\times10^{20}$ $cm^{-3}$, in reasonable agreement with our transport measurement based on a sample synthesized with no extra Te in starting materials (see Figure S3). The combination of the DOS calculations with ARPES and transport measurements suggest that the high Fermi level (~ $E_C$ + 0.2 eV) is obtained under a Te-poor condition. The difference in calculated and measured Fermi levels could result from that the thermodynamic condition assumed in calculations is different from that realized in experiments. We calculated all defect densities at T = 585 °C under thermal equilibrium. Experimentally, different defects freeze into the crystal lattice at different temperatures during the cooling of the crystal because the diffusivities of different defects differ from each other. MnTe has a much higher melting temperature (1150 °C) than $Bi_2Te_3$ does (585 °C), indicating stronger Mn-Te chemical bonds than Bi-Te bonds.

Thus, defects on the Mn sublattice may freeze at a higher temperature than those on the Bi sublattice because defect diffusion requires bonding breaking, which could be more difficult in the MnTe sublattice, resulting in higher defect diffusion barriers. This could lead to a larger concentration difference between $Bi_{Mn}$ and $Mn_{Bi}$ than that calculated at an identical temperature as well as a higher measured Fermi level and higher levels of Mn deficiency than the calculated values.

### D. Defect segregation and surface termination in $MnBi_4Te_7$

$MnBi_4Te_7$ has a more complex structure than $MnBi_2Te_4$ as shown in Figure 1. The separation of magnetic $MnBi_2Te_4$ SLs by nonmagnetic $Bi_2Te_3$ QLs reduces AFM coupling, allowing more freedom in manipulating magnetic and topological properties[10,15,35]. Here, we investigate defect distributions and surface terminations in $MnBi_4Te_7$.

In $MnBi_4Te_7$, the low-energy acceptor defects $Mn_{Bi}^-$ and $Bi_{Te}^-$ can form in either a $MnBi_2Te_4$ SL or a $Bi_2Te_3$ QL. There are two inequivalent sites for $Bi_{Te}^-$ (six- and three-fold coordination) and one site for $Mn_{Bi}^-$ in each SL or QL. The relative energies of $Mn_{Bi}^-$ and $Bi_{Te}^-$ on these different sites are shown in Table II. The formation of $Mn_{Bi}^-$ in the SL is energetically more favorable than in the QL by 0.12 eV. On the other hand, $Bi_{Te}^-$ favors the three-fold coordinated Te site but has little preference on whether this site is in the SL or QL. Therefore, $Bi_{Te}^-$ is distributed in both the SL and QL with nearly the same probability while $Mn_{Bi}^-$ prefers segregation into the QL in $MnBi_4Te_7$. Based on calculated formation energies of $Mn_{Bi}^-$ on different sites, it can be shown that about 5/6 and 1/6 of $Mn_{Bi}^{-}$ are located in the QL and SL, respectively.

Surface energies of $MnBi_2Te_4$- and $Bi_2Te_3$-terminated surfaces in $MnBi_4Te_7$ are calculated (using Eq. 13) to be 0.19 eV and 0.21 eV per surface unit cell. Thus, the $MnBi_2Te_4$-terminated surface is more stable. The magnetic $Mn_{Bi}$ defects are expected to be reduced from this surface due to the gettering effect of $Bi_2Te_3$ QLs as discussed above. However, even 1/6 of the total $Mn_{Bi}^{-}$ population in SLs is significant. At the Te-poor limit, 0.9% of Bi ions are replaced by Mn in SLs, compared to 4.4% replacement in QLs.

Table II. Formation energies of $Mn_{Bi}^{-}$ and $Bi_{Te}^{-}$ on different sites in $MnBi_4Te_7$. There are two sites for $Bi_{Te}^{-}$ (six- and three-fold coordination) and one site (six-fold coordination) for $Mn_{Bi}^{-}$ in each SL $MnBi_2Te_4$ or $Bi_2Te_3$ QL. The energy for the most stable site for each defect is set to zero. The unit is in eV.

| | $MnBi_2Te_4$ SL | | $Bi_2Te_3$ QL | |
|---|---|---|---|---|
| | Six-fold | Three-fold | Six-fold | Three-fold |
| $Mn_{Bi}^{-}$ | 0.12 | N/A | 0 | N/A |
| $Bi_{Te}^{-}$ | 0.75 | < 0.01 | 0.55 | 0 |

## E. Intrinsic defects in $MnSb_2Te_4$

$MnBi_2Te_4$ and $MnBi_4Te_7$ are degenerate n-type semiconductors at the Te-poor limit as shown in Table I. Heavy Sb doping (~30%) has been used to lower the Fermi level, leading to a transition from n- to p-type conductivity in Sb doped $MnBi_2Te_4$.[24-26] We studied intrinsic defects in $MnSb_2Te_4$ (with the AFM ordering) to better understand the effect of Sb doping in $MnBi_2Te_4$. The calculated band gap of AFM $MnSb_2Te_4$ is very small (11 meV), in agreement with a previous study[25]. Thus, we show only the formation energies of the important intrinsic defects at the VBM at both the Te-rich and -poor limits in Table III. There are two important differences between intrinsic defects in $MnSb_2Te_4$ and those in $MnBi_2Te_4$. (1) $MnSb_2Te_4$ favors the formation of acceptor defects ($Mn_{Sb}^{-}$ and $Sb_{Te}^{-}$) over the donor defect $Sb_{Mn}^{+}$, consistent with the transition from n- to p-type conductivity observed in Sb doped $MnBi_2Te_4$.[24] (2) The calculated formation energy of an isolated $Mn_{Sb}^{-}$ is slightly negative, indicating spontaneous formation of $Mn_{Sb}^{-}$ under thermal equilibrium. Therefore, pure $MnSb_2Te_4$ should have a very high defect density and heavy Sb doping of $MnBi_2Te_4$ should increase the concentration of magnetic defects, which can potentially affect the bulk and surface magnetism. The contradictory experimental reports of magnetic ordering in $MnSb_2Te_4$ (AFM in Ref. [24] and FM in Ref. [36]) might be caused by different samples with different $Mn_{Sb}^{-}$ densities and distributions. A high density of $Mn_{Sb}^{-}$ could lead to magnetic ordering of $Mn_{Sb}^{-}$ defects, which mediates a FM interlayer coupling in bulk $MnSb_2Te_4$. [36]

Table III. Calculated formation energies of low-energy intrinsic defects in $MnSb_2Te_4$ at both Te-rich and Te-poor limits [corresponding to points A and B in Figure 2(c)]. The unit is in eV.

| | $Sb_{Mn}^{+}$ | $Mn_{Sb}^{-}$ | $Sb_{Mn}$-$Mn_{Sb}$ | $Sb_{Te}^{-}$ | $V_{Mn}^{2-}$ |
|---|---|---|---|---|---|
| Te-rich | 0.43 | -0.06 | 0.08 | 0.40 | 0.37 |
| Te-poor | 0.37 | -0.01 | 0.08 | 0.08 | 0.50 |

### F. Chemical Doping

As shown in Table I as well as in Figures 3, 4, S3, a Te-rich condition can lower the Fermi level. However, an extremely Te-rich condition close to the Te-rich limit may not be accessible in experiment and the kinetic effect may lead to a higher freeze-in temperature for $\mathrm{Bi}_{\mathrm{Mn}}^{+}$ compared to that of $\mathrm{Mn}_{\mathrm{Bi}}^{-}$, favoring a higher Fermi level as discussed in Sec. II-C. Here, we study extrinsic acceptor dopants, including alkali metal (Li, Na, K) and alkali-earth metal (Be, Mg, Ca) dopants, in $MnBi_2Te_4$ and $MnBi_4Te_7$. We investigated all above dopants in $MnBi_4Te_7$ for its small unit cell, and the most effective acceptor dopant was further studied in $MnBi_2Te_4$.

Among the investigated alkali metal and alkali-earth metal dopants, Na is found to be the most effective p-type dopant. Figures 5 and 6 show calculated formation energies of substitutional $\mathrm{Na}_{\mathrm{Mn}}^{-}$ and $\mathrm{Na}_{\mathrm{Bi}}^{2-}$ as well as interstitial $\mathrm{Na}_{i}^{+}$ in $MnBi_2Te_4$ and $MnBi_4Te_7$. It can be seen that $\mathrm{Na}_{\mathrm{Mn}}^{-}$ has a lower formation energy than $\mathrm{Na}_{\mathrm{Bi}}^{2-}$, $\mathrm{Na}_{i}^{+}$ , and the most stable intrinsic acceptor defect; therefore, $\mathrm{Na}_{\mathrm{Mn}}^{-}$ can act as the dominant acceptor, which compensates the intrinsic donor defect $\mathrm{Bi}_{\mathrm{Mn}}^{+}$, at both Te-rich and -poor limits in both $MnBi_2Te_4$ and $MnBi_4Te_7$.

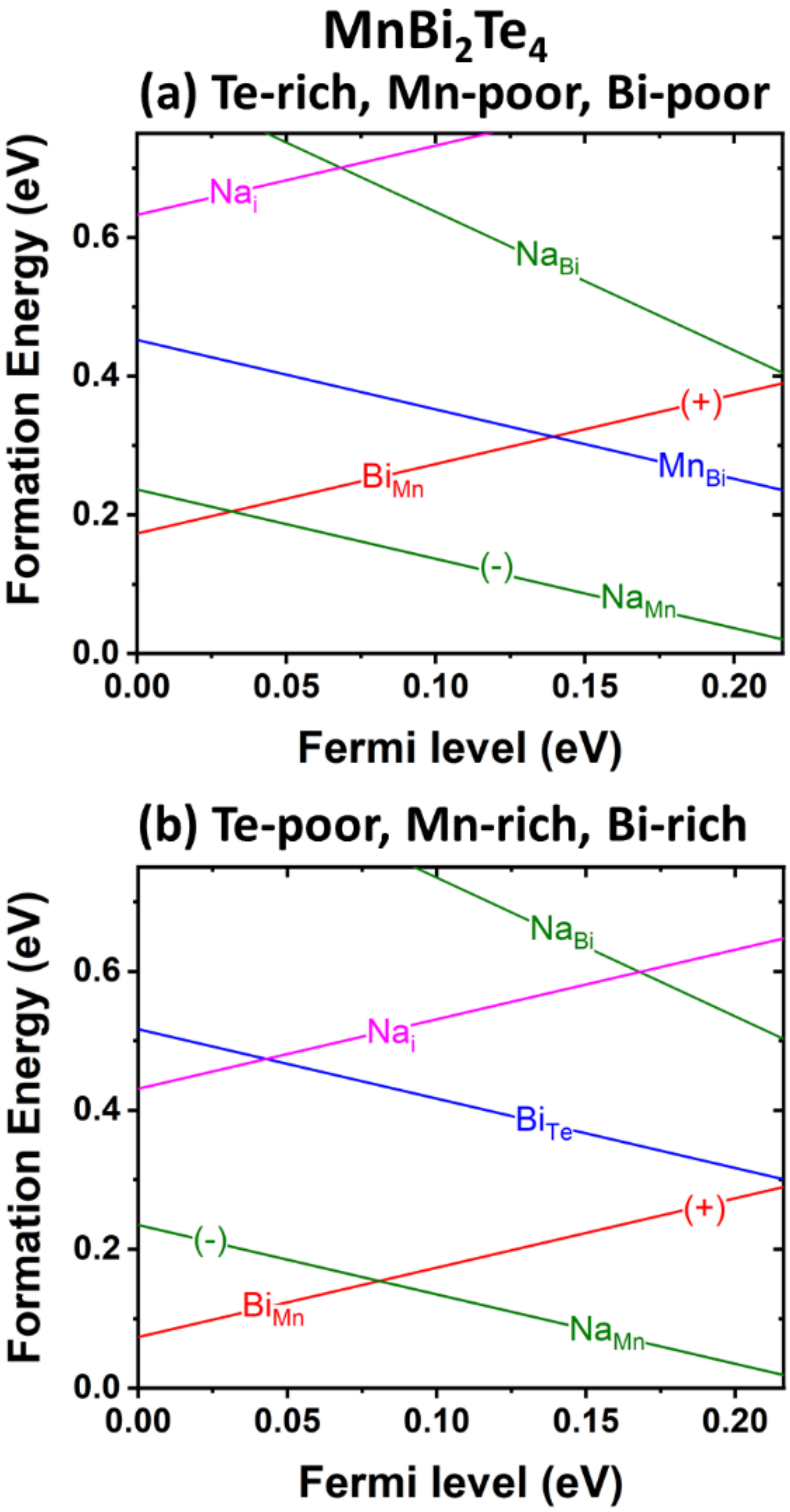

Figure 5. Calculated formation energies of $Na_{Mn}^{-}$, $Na_{Bi}^{2-}$, $Na_{i}^{+}$, and the most stable intrinsic defects as functions of the Fermi level (varied from the VBM to the CBM) at the Te-rich/cation-poor (a) and Te-poor/cation-rich (b) limits [corresponding to points A and B in Figure 2(a), respectively] in $MnBi_2Te_4$. The Na-rich limit is applied in both (a) and (b). The slope of a formation energy line indicates the charge state of the defect as selectively shown.

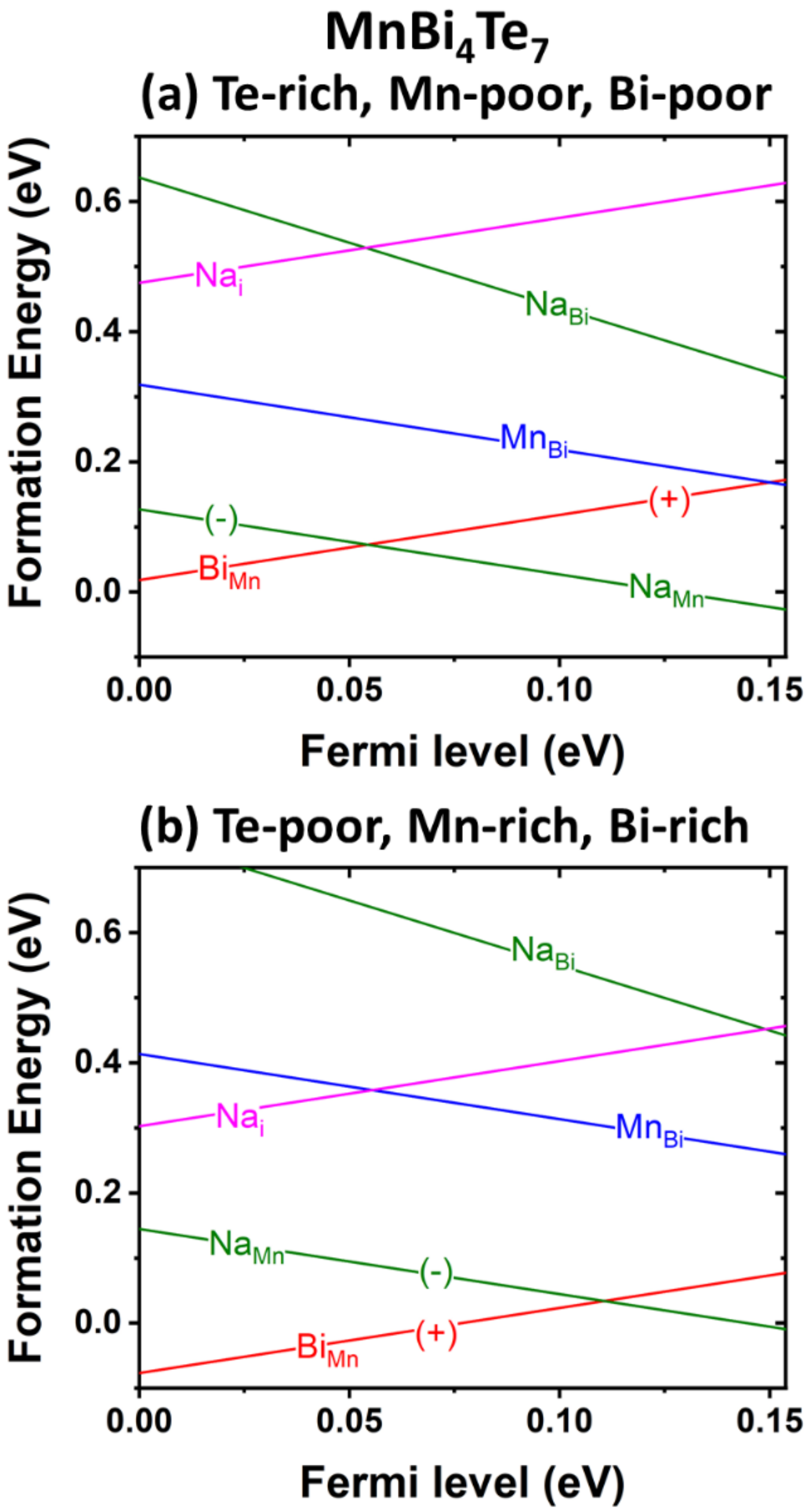

Figure 6. Calculated formation energies of $Na_{Mn}^{-}$, $Na_{Bi}^{2-}$, $Na_{i}^{+}$, and the most stable intrinsic defects as functions of the Fermi level (varied from the VBM to the CBM) at the Te-rich/cation-poor (a) and Te-poor/cation-rich (b) limits [corresponding to points A and B in Figure 2(b), respectively] in $MnBi_4Te_7$. The Na-rich limit is applied in both (a) and (b). The slope of a formation energy line indicates the charge state of the defect as selectively shown.

As shown in Table I, both $MnBi_2Te_4$ and $MnBi_4Te_7$ can be tuned p-type at the Te-rich limits. However, the experimental growth condition appears to be close to the Te-poor limit, resulting in the Fermi level above the CBM as discussed in Sec. II-C. Thus, we focus on Na doping at the Te-poor limit. Formation energies of Na dopants shown in Figures 5-6 are the lowest values calculated at the Na-rich limit [highest Na chemical potential allowed by Eq. (8)]. At the Na-rich and Te-poor limits, the calculated [$Na_{Mn}^-$] and [$Bi_{Mn}^+$] are $5.7\times10^{20}$ cm$^{-3}$ (12.6%) and $5.5\times10^{20}$ cm$^{-3}$ (12.3%), respectively, in $MnBi_2Te_4$ and $1.5\times10^{21}$ cm$^{-3}$ (59.7%) and $1.7\times10^{21}$ cm$^{-3}$ (66.9%), respectively, in $MnBi_4Te_7$. These are very high densities. In the case of $MnBi_4Te_7$, the calculated defect and dopant densities are unrealistic; thus, calculations of the formation energy of an isolated defect/dopant [Eq. (8)] and its density [Eq. (9)] cannot be applied to the case of high defect/dopant densities at the Na-rich/Te-poor limit.

To better understand the effect of Na doping, we studied a range of lower Na doping levels below the solid solubility of Na, and calculated densities of free carriers, intrinsic defects and the Fermi level as functions of [$Na_{Mn}^-$] at the Te-poor limit in both $MnBi_2Te_4$ and $MnBi_4Te_7$ (see Figure 7). In $MnBi_2Te_4$, 1.6% of $Na_{Mn}^-$ doping ($7\times10^{19}$ cm$^{-3}$) can position the Fermi level near the midgap ($E_V$ + 0.09 eV), at which [$Bi_{Mn}^+$] is about 4.2% ($1.88\times10^{19}$ cm$^{-3}$) [see Figure 7(a)]. On the other hand, the Fermi level in $MnBi_4Te_7$ remains above the CBM despite heavy Na doping [up to nearly 40% ($1\times10^{21}$ cm$^{-3}$)] as shown in Figure 7(b). Therefore, Na doping can lead to the insulating behavior in $MnBi_2Te_4$ even at the Te-poor limit but must be combined with a more Te-rich growth condition to lower the Fermi level into the band gap in $MnBi_4Te_7$.

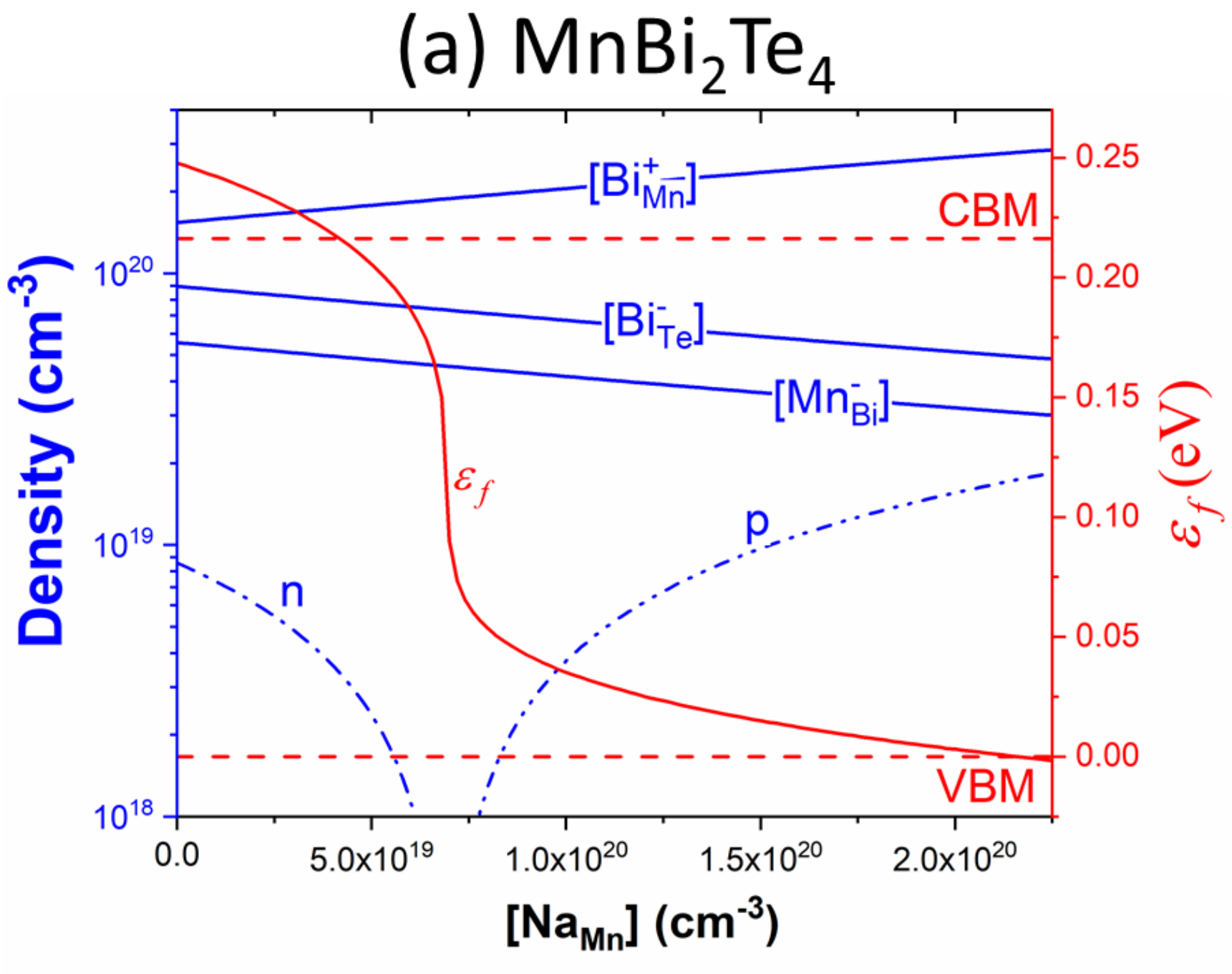

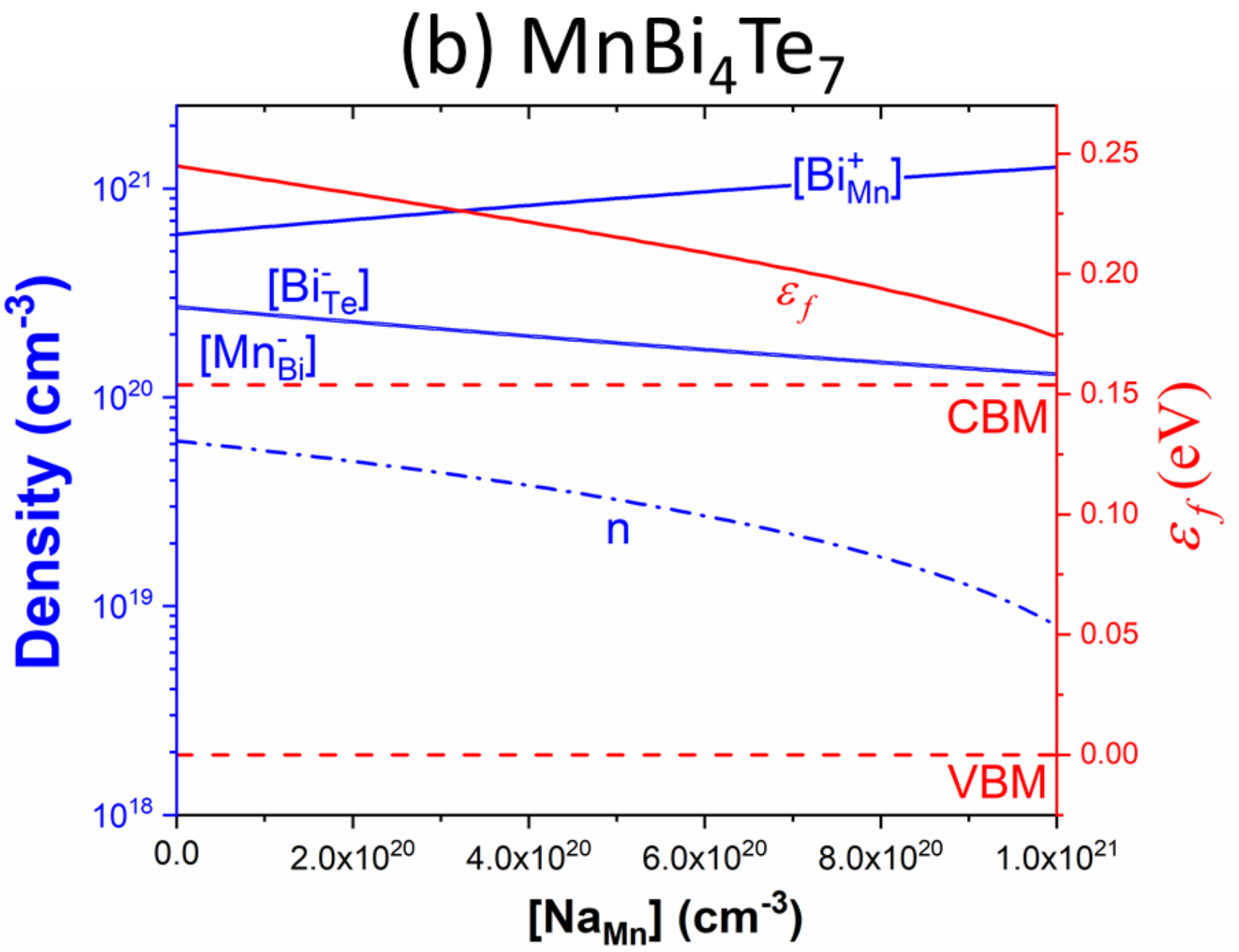

Figure 7. Densities of free electrons (n) and holes (p) and most important intrinsic defects (in blue) as well as the Fermi level (in red) as functions of the substitutional $Na_{Mn}$ density calculated at the Te-poor limit in (a) $MnBi_2Te_4$ and (b) $MnBi_4Te_7$. The densities of native Mn ions are $4.49\times10^{21}$ cm$^{-3}$ and $2.53\times10^{21}$ cm$^{-3}$ in $MnBi_2Te_4$ and $MnBi_4Te_7$, respectively.

# III. Discussion

### A. Large-size-mismatched antisite defects and substitutional dopants

Our defect calculations show that $Bi_{Mn}$ has a very low formation energy in $MnBi_2Te_4$ and $MnBi_4Te_7$ especially in the latter (Figures 3-4) despite that the ionic radius of $Bi^{3+}$ (1.03 Å) is much larger than that of $Mn^{2+}$ (0.83 Å).[37] The stability of such large-sized-mismatched antisite defect is likely related to strain in the MnTe layer. The MnTe layer is intercalated within the $Bi_2Te_3$ layer in a SL, forming a natural internal heterostructure. The calculated in-plane lattice constants of hexagonal lattices of MnTe, $MnBi_2Te_4$, $MnBi_4Te_7$, and $Bi_2Te_3$ increase from 4.160 Å, 4.365 Å, 4.394 Å to 4.433 Å. These results suggest that the MnTe layer is under significant tensile strain in $MnBi_2Te_4$. Such strain is further increased in $MnBi_4Te_7$ because an additional $Bi_2Te_3$ QL in the unit cell of $MnBi_4Te_7$ expands the in-plane lattice constant. Therefore, substitution of Mn by a large ion reduces the tensile strain in the MnTe layer. This could explain the low formation energy of $Bi_{Mn}$. Opposite to the tensile strain in the inner MnTe layer, the outer $Bi_2Te_3$ layer that encapsulates the MnTe layer is under the compressive strain, which promotes the substitution of Bi by a smaller Mn ion. However, the $Bi_2Te_3$ layer is thicker and adjacent to the vdW gap; thus, the strain in the $Bi_2Te_3$ layer is better relaxed than that in the MnTe layer. This is consistent with the generally lower formation energy of $Bi_{Mn}$ than that of $Mn_{Bi}$. Compared to $MnBi_2Te_4$, the tensile strain in the MnTe layer is larger and the compressive strain in the $Bi_2Te_3$ layer is smaller in $MnBi_4Te_7$, resulting in a lower $Bi_{Mn}$ formation energy and a higher $Mn_{Bi}$ formation energy (hence, increased Mn deficiency) in $MnBi_4Te_7$. Based on the above analysis, it may be expected that $MnBi_6Te_{10}$, which has two $Bi_2Te_3$ QLs in the unit cell, should have more $Bi_{Mn}$ and less $Mn_{Bi}$ than $MnBi_4Te_7$ and $MnBi_2Te_4$. Indeed, previous experiments showed an increased Mn deficiency in $MnBi_6Te_{10}$.[21,30]

The strain effect described above also determines the doping efficiency. Substitutional doping usually favors dopants with sizes close to that of the substituted native atom. However, due to the tensile strain in the MnTe layer, a large-sized dopant with the size close to that of Bi is favored on the Mn site. Similarly, due to the compressive strain in the $Bi_2Te_3$ layer, a small-sized dopant with the size close to that of Mn is favored on the Bi site. This is demonstrated by alkali-metal doping in $MnBi_4Te_7$ as shown in Figure 8. $Li^+$ has an ionic radius of 0.76 Å, close to that of $Mn^{2+}$ (0.83 Å) while $Na^+$ has an ionic radius of 1.02 Å, close to that of $Bi^{3+}$ (1.03 Å).[37] Nevertheless, the large-size-mismatched $\mathrm{Na_{Mn}^-}$ has a lower formation energy than $\mathrm{Li_{Mn}^-}$. $\mathrm{K_{Mn}^-}$ has a very high formation energy because $K^+$ has an ionic radius of 1.38 Å, even much larger than the size of $Bi^{3+}$. Turning to alkali-earth-metal dopants, the size of $Ca^{2+}$ (1 Å) is close to that of $Bi^{3+}$ (1.03 Å) and the size of $Mg^{2+}$ (0.72 Å) is close to that of $Mn^{2+}$ (0.83 Å).[37] Again, the large-size-mismatched $\mathrm{Mg_{Bi}^-}$ has a lower formation energy than $\mathrm{Ca_{Bi}^-}$ in $MnBi_4Te_7$ as shown in Figure 9. We have also tested $\mathrm{Be_{Bi}^-}$. The small ionic radius of $Be^{2+}$ (0.45 Å) causes the off centering of the Be ion on the Bi site, leading to a very high formation energy (not shown). Figure 9 also shows that alkali-earth-metal dopants favor the doping on the isovalent Mn site, rendering them electrically inactive and ineffective in tuning the Fermi level.

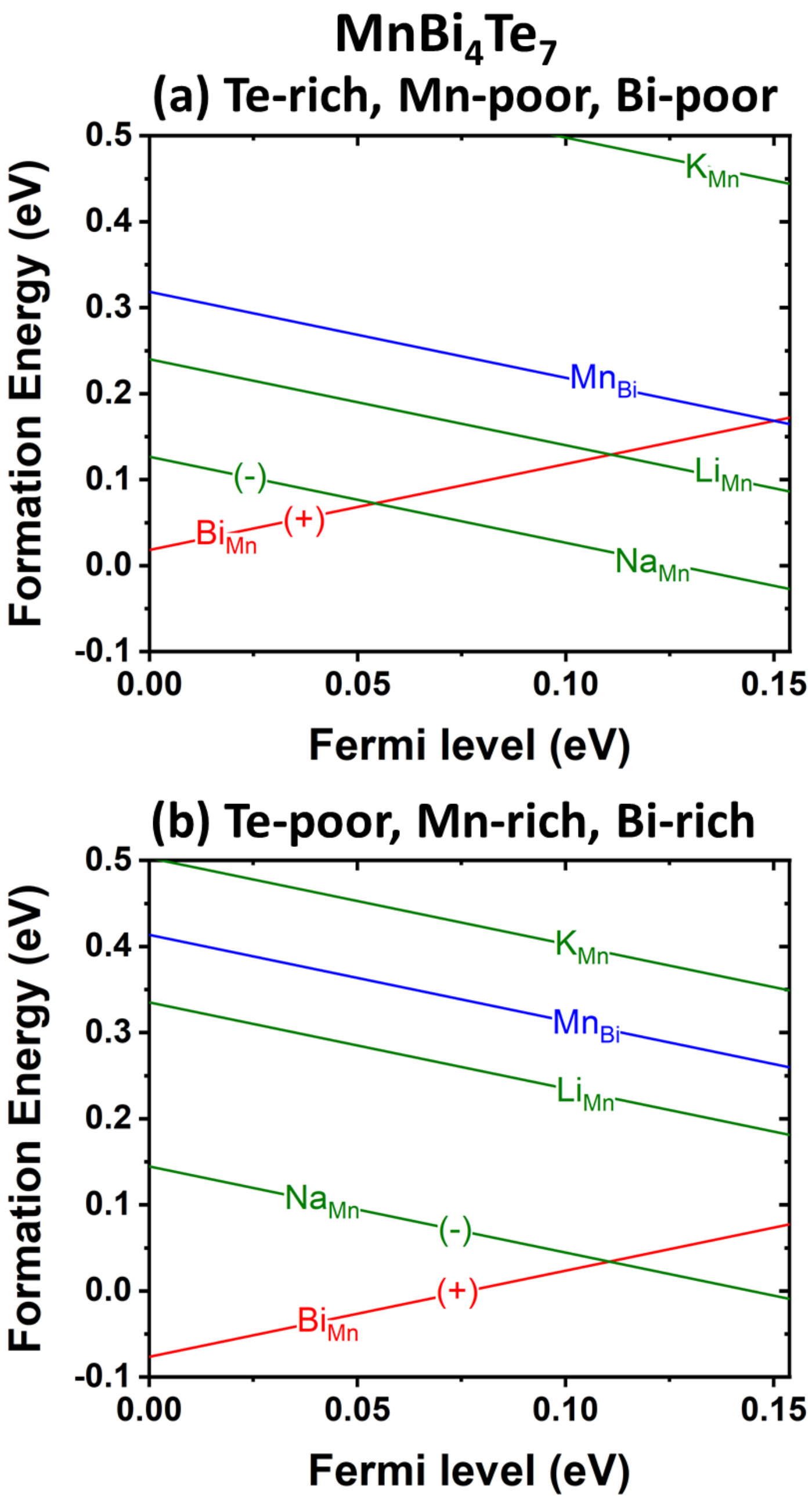

Figure 8. Calculated formation energies of $Li_{Mn}^{-}$, $Na_{Mn}^{-}$, $K_{Mn}^{-}$, and the most stable intrinsic defects as functions of the Fermi level (varied from the VBM to the CBM) at the Te-rich/cation-poor (a) and Te-poor/cation-rich (b) limits [corresponding to points A and B in Figure 2(b), respectively] in $MnBi_4Te_7$. The dopant-rich limit is applied in both (a) and (b). The slope of a formation energy line indicates the charge state of the defect as selectively shown.

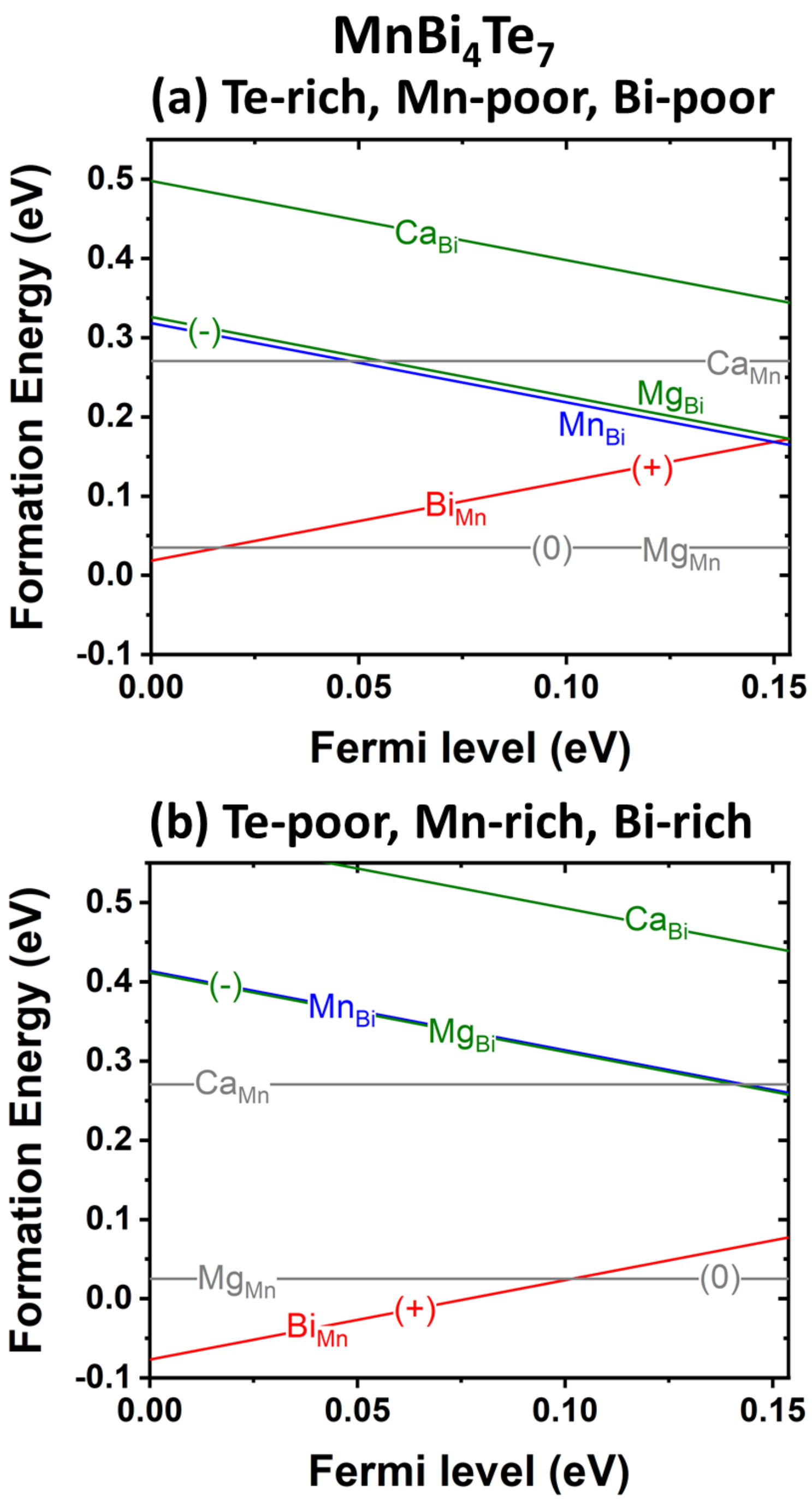

Figure 9. Calculated formation energies of substitutional Mg and Ca dopants and the most stable intrinsic defects as functions of the Fermi level (varied from the VBM to the CBM) at the Te-rich/cation-poor (a) and Te-poor/cation-rich (b) limits [corresponding to points A and B in Figure 2(b), respectively] in $MnBi_4Te_7$. The dopant-rich limit is applied in both (a) and (b). The slope of a formation energy line indicates the charge state of the defect as selectively shown.

### B. Composition analysis

Our defect calculations shown in Sec. II-B and C suggest that several assumptions made in the experimental composition analysis [14,21,30] could lead to significant errors. (1) The charge state of Bi ions was assumed to be +3 when the charge neutrality condition was applied in the composition analysis. However, the Bi ion in $\mathrm{Bi}_{\mathrm{Te}}^{-}$ has a charge state of -1 and $\mathrm{Bi}_{\mathrm{Te}}^{-}$ is abundant and has a density comparable to that of $\mathrm{Mn}_{\mathrm{Bi}}^{-}$ in both $MnBi_2Te_4$ and $MiBi_4Te_7$ at the Te-poor limit (Table I). (2) Previous composition analyses did not consider the free carrier density when applying the charge neutrality condition. However, the free electron density can be on the order of $10^{20}$ $cm^{-3}$ (Figure S3), comparable to defect densities. (3) The Mn vacancy, $V_{\mathrm{Mn}}^{2-}$, which was assumed as an important defect in previous composition analyses, actually has a relatively high formation energy in both $MnBi_2Te_4$ and $MiBi_4Te_7$ (Figures 3-4) and should not be a major factor behind the experimentally observed Mn deficiency in $MnBi_4Te_7$ [14,21,30]. The neglect of the negatively charged $\mathrm{Bi}_{\mathrm{Te}}^{-}$ and free electrons could increase the estimated density of positively charged defects ($\mathrm{Bi}_{\mathrm{Mn}}^{+}$), leading to artificially high Mn deficiencies in composition analyses.

## IV. Conclusion

We show that the intercalation of the MnTe layer within the $Bi_2Te_3$ layer in antiferromagnetic topological insulators $MnBi_2Te_4$ and $MnBi_4Te_7$ creates an internal heterostructure with large strain, promoting the formation of large-size-mismatched antisite defect $\mathrm{Bi}_{\mathrm{Mn}}$, which is shown to be more abundant in $MnBi_4Te_7$ than in $MnBi_2Te_4$ due to the larger tensile strain in the MnTe layer in the former. The abundance of the intrinsic donor defect $Bi_{Mn}$ positions the Fermi level above the CBM under the Te-poor growth condition, giving rise to n-type metallic conductivity. Our DFT calculations and preliminary synthesis demonstrate that adopting a Te-rich condition can lower the Fermi level, which may be pinned inside the bulk energy gap of $MnBi_2Te_4$ and $MnBi_4Te_7$ – a condition needed for quantum transport measurement of surface states. In contrast, the intrinsic acceptor defect $Mn_{Sb}$ is dominant in $MnSb_2Te_4$, leading to p-type conductivity. The strain at the internal heterostructure also enables the efficient incorporation of large-size-mismatched substitutional acceptor dopant $Na_{Mn}$, which can have a density higher than any intrinsic defects. We show that $Na_{Mn}$ can compensate $Bi_{Mn}$ and pin the Fermi level inside the band gap even at the Te-poor limit in $MnBi_2Te_4$. The high density of magnetic defects ($Mn_{Bi}$ in $MnBi_2Te_4$ and $MnBi_4Te_7$ and $Mn_{Sb}$ in $MnSb_2Te_4$) found in this study may have important implications to magnetic and topological properties and deserve further studies.

## V. Methods

### A. Defect and dopant formation energy calculations

The formation energy of a defect or a dopant is given by

$$\Delta H(q,\varepsilon_f) = (E_D - E_h) - \sum_i n_i\left(\mu_i + \mu_i^{bulk}\right) + q(\varepsilon_{VBM} + \varepsilon_f), \qquad (1)$$

where $E_D$ and $E_h$ are the total energies of the defect-containing and the host (i.e. defect-free) supercells. The formation of a defect in a material involves an exchange of atoms with their respective chemical reservoirs. The second term in Eq. (1) represents the change in energy due to such exchange of atoms, where $n_i$ is the difference in the number of atoms for the $i$'th atomic species between the defect-containing and defect-free supercells. $\mu_i$ is the relative chemical potential for the $i$'th atomic species, referenced to the chemical potential of its elemental bulk phase $\mu_i^{bulk}$. The third term in Eq. (1) represents the change in energy due to the exchange of electrons with its reservoir. $q$ is the charge state of the defect. $\varepsilon_{VBM}$ is the energy of the VBM and $\varepsilon_f$ is the Fermi energy relative to the VBM. The correction to the defect formation energy due to potential alignment (between the host and a charged defect supercell) [38] was applied. The image charge correction was not included because the supercell size along the c axis and the static dielectric constant on the a-b plane are very large, suppressing the image charge interaction. [The supercell sizes along c are 80.951 Å for $MnBi_2Te_4$, 47.267 Å for $MnBi_4Te_7$, and 80.601 Å for $MnSb_2Te_4$. The calculated static dielectric constant on the a-b plane of $MnBi_4Te_7$ is 225 (the electronic and ionic contributions are 107 and 118, respectively.) The dielectric constants of $MnBi_2Te_4$ and $MnSb_2Te_4$ are expected to be large as well.]

The chemical potentials in Eq. (1) are subject to a series of thermodynamic constraints under the equilibrium growth condition. To maintain the stability of $MnBi_2Te_4$ during growth, the chemical potentials of Mn, Bi, and Te should satisfy

$$\mu_{\mathrm{Mn}} + 2\mu_{\mathrm{Bi}} + 4\mu_{\mathrm{Te}} = \Delta H(\mathrm{MnBi_2Te_4}), \qquad (2)$$

where $\Delta H(\mathrm{MnBi_2Te_4})$ is the enthalpy of formation for $MnBi_2Te_4$. Eq. 2 reduces the number of independent elemental chemical potentials to two. We chose $\mu_{\mathrm{Bi}}$ and $\mu_{\mathrm{Te}}$ as the two independent elemental chemical potentials to plot Figure 2. $\mu_{\mathrm{Mn}}$ can be determined by Eq. 2.

To avoid the formation of binary phases (MnTe, $MnTe_2$, $Bi_2Te_3$, BiTe, $Bi_4Te_3$, $Bi_8Te_9$) and elemental phases of Mn, Bi, and Te, the following constraints on chemical potentials are applied:

$$\begin{aligned}
&\mu_{\mathrm{Mn}} + \mu_{\mathrm{Te}} \leq \Delta H(\mathrm{MnTe}),\\
&\mu_{\mathrm{Mn}} + 2\mu_{\mathrm{Te}} \leq \Delta H(\mathrm{MnTe_2}),\\
&2\mu_{\mathrm{Bi}} + 3\mu_{\mathrm{Te}} \leq \Delta H(\mathrm{Bi_2Te_3}),\\
&\mu_{\mathrm{Bi}} + \mu_{\mathrm{Te}} \leq \Delta H(\mathrm{BiTe}),\\
&4\mu_{\mathrm{Bi}} + 3\mu_{\mathrm{Te}} \leq \Delta H(\mathrm{Bi_4Te_3}),\\
&8\mu_{\mathrm{Bi}} + 9\mu_{\mathrm{Te}} \leq \Delta H(\mathrm{Bi_8Te_9}),\\
&\mu_{\mathrm{Mn}} \leq 0,\ \mu_{\mathrm{Bi}} \leq 0,\ \mu_{\mathrm{Te}} \leq 0.
\end{aligned} \tag{3}$$

Here, $\Delta H(\mathrm{MnTe})$, $\Delta H(\mathrm{MnTe_2})$, $\Delta H(\mathrm{Bi_2Te_3})$, $\Delta H(\mathrm{BiTe})$, $\Delta H(\mathrm{Bi_4Te_3})$, and $\Delta H(\mathrm{Bi_8Te_9})$ are enthalpies of formation for MnTe, $MnTe_2$, $Bi_2Te_3$, BiTe, $Bi_4Te_3$, and $Bi_8Te_9$, respectively.

For the growth of $MnBi_4Te_7$, the chemical potentials of Mn, Bi, and Te should satisfy

$$\mu_{\mathrm{Mn}} + 4\mu_{\mathrm{Bi}} + 7\mu_{\mathrm{Te}} = \Delta H(\mathrm{MnBi_4Te_7}), \tag{4}$$

where $\Delta H(\mathrm{MnBi_4Te_7})$ is the enthalpy of formation for $MnBi_4Te_7$. All chemical potential constraints in Eq. (3) also apply to $MnBi_4Te_7$ with an additional constraint for avoiding the formation of $MnBi_2Te_4$:

$$\mu_{\mathrm{Mn}} + 2\mu_{\mathrm{Bi}} + 4\mu_{\mathrm{Te}} \leq \Delta H(\mathrm{MnBi_2Te_4}). \tag{5}$$

For the growth of $MnSb_2Te_4$, the chemical potentials of Mn, Sb, and Te should

satisfy

$$\mu_{\mathrm{Mn}} + 2\mu_{\mathrm{Sb}} + 4\mu_{\mathrm{Te}} = \Delta H(\mathrm{MnSb_2Te_4}), \quad (6)$$

where $\Delta H(\mathrm{MnSb_2Te_4})$ is the enthalpy of formation for $MnSb_2Te_4$. The following constraints are applied to avoid the formation of secondary phases (MnTe, $MnTe_2$, $Sb_2Te_3$, SbTe, $Sb_2Te$, $Sb_8Te_3$, Mn, Sb, Te):

$$\begin{aligned} &\mu_{\mathrm{Mn}} + \mu_{\mathrm{Te}} \le \Delta H(\mathrm{MnTe}), \\ &\mu_{\mathrm{Mn}} + 2\mu_{\mathrm{Te}} \le \Delta H(\mathrm{MnTe_2}), \\ &2\mu_{\mathrm{Sb}} + 3\mu_{\mathrm{Te}} \le \Delta H(\mathrm{Sb_2Te_3}), \\ &\mu_{\mathrm{Sb}} + \mu_{\mathrm{Te}} \le \Delta H(\mathrm{SbTe}), \\ &2\mu_{\mathrm{Sb}} + \mu_{\mathrm{Te}} \le \Delta H(\mathrm{Sb_2Te}), \\ &8\mu_{\mathrm{Sb}} + 3\mu_{\mathrm{Te}} \le \Delta H(\mathrm{Sb_8Te_3}), \\ &\mu_{\mathrm{Mn}} \le 0,\ \mu_{\mathrm{Sb}} \le 0,\ \mu_{\mathrm{Te}} \le 0. \end{aligned} \quad (7)$$

Here, $\Delta H(\mathrm{Sb_2Te_3})$, $\Delta H(\mathrm{SbTe})$, $\Delta H(\mathrm{Sb_2Te})$, and $\Delta H(\mathrm{Sb_8Te_3})$ are enthalpies of formation for $Sb_2Te_3$, SbTe, $Sb_2Te$, and $Sb_8Te_3$, respectively.

After considering all above constraints, we find that the stable phase of $MnBi_2Te_4$ is confined by MnTe, $MnTe_2$, $Bi_2Te_3$, and BiTe phases [Figure 2(a)]; the stable phase of $MnBi_4Te_7$ is confined by MnTe, $MnTe_2$, $Bi_2Te_3$, BiTe, and $MnBi_2Te_4$ phases [Figure 2(b)]; the stable phase of $MnSb_2Te_4$ is confined by MnTe, $MnTe_2$, $Sb_2Te_3$, and SbTe phases [Figure 2(c)]. Other phases considered in Eqs (3) and (7) do not share boundaries with the targeted ternary phase and thus are not shown in Figure 2. It is shown in Sec. II-A that $MnBi_2Te_4$ is marginably stable against decomposition to MnTe and $Bi_2Te_3$ whereas $MnBi_4Te_7$ and $MnSb_2Te_4$ are slightly metastable compared to binary phases (all by a few meV, close to the numerical uncertainty). To simplify the problem, we make following approximations: $\Delta H(\mathrm{MnTe}) + \Delta H(\mathrm{Bi_2Te_3}) = \Delta H(\mathrm{MnBi_2Te_4})$, $\Delta H(\mathrm{MnTe}) + 2\Delta H(\mathrm{Bi_2Te_3}) = \Delta H(\mathrm{MnBi_4Te_7})$, $\Delta H(\mathrm{MnBi_2Te_4}) + \Delta H(\mathrm{Bi_2Te_3}) = \Delta H(\mathrm{MnBi_4Te_7})$, and $\Delta H(\mathrm{MnTe}) + \Delta H(\mathrm{Sb_2Te_3}) = \Delta H(\mathrm{MnSb_2Te_4})$. As a result, the chemical potential ranges in $MnBi_2Te_4$, $MnBi_4Te_7$ and $MnSb_2Te_4$ are line segments as shown in the phase diagrams of Figure 2, which introduces errors of a few meV.

Chemical doping of $MnBi_2Te_4$ and $MnBi_4Te_7$ by Li, Na, K, Be, Mg, and Ca are limited by the formation of dopant-Te secondary phases. The following constraints are, thus, applied:

$$
\begin{aligned}
&2\mu_{\mathrm{Li}} + \mu_{\mathrm{Te}} \le \Delta H(\mathrm{Li_2Te}),\\
&\mu_{\mathrm{Li}} + 3\mu_{\mathrm{Te}} \le \Delta H(\mathrm{LiTe_3}),\\
&2\mu_{\mathrm{Na}} + \mu_{\mathrm{Te}} \le \Delta H(\mathrm{Na_2Te}),\\
&\mu_{\mathrm{Na}} + 3\mu_{\mathrm{Te}} \le \Delta H(\mathrm{NaTe_3}),\\
&2\mu_{\mathrm{K}} + \mu_{\mathrm{Te}} \le \Delta H(\mathrm{K_2Te}),\\
&\mu_{\mathrm{K}} + \mu_{\mathrm{Te}} \le \Delta H(\mathrm{KTe}),\\
&2\mu_{\mathrm{K}} + 3\mu_{\mathrm{Te}} \le \Delta H(\mathrm{K_2Te_3}),\\
&5\mu_{\mathrm{K}} + 3\mu_{\mathrm{Te}} \le \Delta H(\mathrm{K_5Te_3}),\\
&\mu_{\mathrm{Be}} + \mu_{\mathrm{Te}} \le \Delta H(\mathrm{BeTe}),\\
&\mu_{\mathrm{Mg}} + \mu_{\mathrm{Te}} \le \Delta H(\mathrm{MgTe}),\\
&\mu_{\mathrm{Mg}} + 2\mu_{\mathrm{Te}} \le \Delta H(\mathrm{MgTe_2})\\
&\mu_{\mathrm{Ca}} + \mu_{\mathrm{Te}} \le \Delta H(\mathrm{CaTe}).
\end{aligned}
\tag{8}
$$

Here, $\Delta H(\mathrm{Li_2Te})$, $\Delta H(\mathrm{LiTe_3})$, $\Delta H(\mathrm{Na_2Te})$, $\Delta H(\mathrm{NaTe_3})$, $\Delta H(\mathrm{K_2Te})$, $\Delta H(\mathrm{KTe})$,

$\Delta H(\mathrm{K_2Te_3})$, $\Delta H(\mathrm{K_5Te_3})$, $\Delta H(\mathrm{BeTe})$, $\Delta H(\mathrm{MgTe})$, $\Delta H(\mathrm{MgTe_2})$, and $\Delta H(\mathrm{CaTe})$ are enthalpies of formation for $Li_2Te$, $LiTe_3$, $Na_2Te$, $NaTe_3$, $K_2Te$, KTe, $K_2Te_3$, $K_5Te_3$, BeTe, MgTe, and CaTe, respectively. Chemical potentials of Li, Na, K, Be, Mg, and Ca in the growth of $MnBi_4Te_7$ are capped by the formation of $Li_2Te$, $NaTe_3$, $K_2Te_3$, BeTe, $MgTe_2$, and CaTe, respectively, at the Te-rich limit and by the formation of $Li_2Te$, $Na_2Te$, $K_2Te_3$, BeTe, MgTe, and CaTe at the Te-poor limit. The chemical potential of Na in the growth of $MnBi_2Te_4$ is capped by the formation of $NaTe_3$ at the Te-rich limit and by the formation of $Na_2Te$ at the Te-poor limit.

The defect density under thermal equilibrium can be calculated by

$$N_D(q,\varepsilon_f) = N_{\mathrm{site}} e^{\frac{-\Delta H(q,\varepsilon_f)}{k_B T}}, \tag{9}$$

where $N_{\mathrm{site}}$ is the number of available atomic sites for defect formation, $\Delta H(q,\varepsilon_f)$ is the defect formation energy calculated by Eq. (1), $k_B$ is the Boltzmann constant, and T is temperature. The Fermi level, $\varepsilon_f$, is determined by solving the following equation to satisfy the charge neutrality condition:

$$\sum_{i,j} q_i N_{D_j}(q_i,\varepsilon_f) + n_h - n_e = 0. \tag{10}$$

Here, $N_{D_j}(q_i)$ is the number of the $j$th defect with a charge state of $q_i$, calculated using the Eq. 9. The free hole ($n_h$) and free electron ($n_e$) densities are calculated by

$$n_h = \int_{-\infty}^{\varepsilon_{VBM}} N_{DOS}(E)\left[1 - f_e\left(E,\varepsilon_f,T\right)\right]dE \tag{11}$$

and

$$n_e = \int_{\varepsilon_{CBM}}^{\infty} N_{DOS}(E) f_e\left(E,\varepsilon_f,T\right)dE, \tag{12}$$

respectively. $N_{DOS}(E)$ is the calculated density of states and $f_e\left(E,\varepsilon_f\right)$ is the Fermi-Dirac distribution.

## B. Surface energy calculations

We constructed a $MnBi_2Te_4$- and a $Bi_2Te_3$-terminated symmetric slabs for calculations of surface energies of the two different terminations of the $MnBi_4Te_7$ (0001) surface. The $MnBi_2Te_4$-terminated symmetric slab contains five $MnBi_2Te_4$ SLs and four $Bi_2Te_3$ QLs. The $Bi_2Te_3$-terminated symmetric slab contains six $Bi_2Te_3$ QLs and five $MnBi_2Te_4$ SLs. The vacuum layer included for slab calculations has a thickness of 16 Å. We also tested a much thinner $MnBi_2Te_4$-terminated symmetric slab, which contains only three $MnBi_2Te_4$ SLs and two $Bi_2Te_3$ QLs. The calculated surface energy difference is only 1.5 meV per surface unit cell. The surface energy is calculated by

$$\Delta H_{surf} = [E_{slab} - \sum_i n_i(\mu_i + \mu_i^{bulk})]/2A. \quad (13)$$

Here, $E_{slab}$ is the total energy of the slab; $\mu_i$ and $\mu_i^{bulk}$ are same as those in Eq. (1); A is the surface area. Within the allowed chemical range [the line segment between points A and B in Fig. 2(b)], the following relations always hold: $\mu_{\mathrm{Mn}} + 2\mu_{\mathrm{Bi}} + 4\mu_{\mathrm{Te}} = \Delta H(\mathrm{MnBi_2Te_4})$ and $2\mu_{\mathrm{Bi}} + 3\mu_{\mathrm{Te}} = \Delta H(\mathrm{Bi_2Te_3})$. Based on these, it can be shown that the calculated surface energy is independent of elemental chemical potentials.

## C. Computational details

All calculations are based on density functional theory (DFT)[39,40] implemented in the VASP code.[41] The interaction between ions and electrons is described by the projector augmented wave method[42]. The total energy is calculated using the Perdew-Burke-Eznerhof (PBE) exchange correlation functional[43] and a kinetic energy cutoff of 270 eV. A $U$ parameter of 4 eV is applied to Mn 3d orbitals[44] and the DFT-D3 vdW functional [45] is used, following several previous DFT studies.[11,12,17] A 3×3×2 supercell and a 2×2×1 k-point mesh are used for defect calculations. The c axis is doubled for the AFM calculation. Six SLs are included for $MnBi_2Te_4$ and $MnSb_2Te_4$ while two SLs and two QLs are included for $MnBi_4Te_7$. Lattice parameters were optimized, and atomic positions were relaxed until the forces are less than 0.02 eV/Å. The optimized lattice constants for MnTe, $Bi_2Te_3$, $Sb_2Te_3$, $MnBi_2Te_4$, and $MnBi_4Te_7$ are in good agreement with experimental values as shown in Table IV.

Table IV. Calculated lattice parameters compared with experimentally measured values.

| | | MnTe | $Bi_2Te_3$ | $Sb_2Te_3$ | $MnBi_2Te_4$ | $MnBi_4Te_7$ | $MnSb_2Te_4$ |
|---|---|---|---|---|---|---|---|
| Cal. | a (Å) | 4.158 | 4.433 | 4.325 | 4.365 | 4.394 | 4.289 |
| | c (Å) | 6.703 | 30.392 | 29.973 | 40.476 | 23.634 | 40.301 |
| Exp. | a (Å) | 4.148 | 4.395 | 4.264 | 4.3338 | 4.366 | 4.2445 |
| | c (Å) | 6.711 | 30.44 | 30.458 | 40.931 | 23.80 | 40.870 |
| | | [46] | [47] | [48] | [24] | [21] | [24] |

As mentioned above, a $U$ parameter of 4 eV was used. We tested $U$ = 3 eV, 4 eV, and 5 eV for the formation energy calculation of the antisite pair $Bi_{Mn}$-$Mn_{Bi}$ in $MnBi_4Te_7$, which yields 0.34 eV, 0.33 eV, and 0.32 eV, respectively. This result shows that different choices of the $U$ parameter have a small influence on the defect formation energy.

$U$ = 4 eV is appropriate for an insulating Mn compound with local Mn moments, such as $MnBi_2Te_4$ (calculated magnetic moments: 4.54 $\mu_B$ in $MnBi_2Te_4$ and $MnBi_4Te_7$ and 4.52 $\mu_B$ in $MnSb_2Te_4$) but should not be applied to Mn metal, which has an itinerant nature. However, for calculations of enthalpies of formation for $MnBi_2Te_4$, $MnBi_4Te_7$, $MnSb_2Te_4$, MnTe, and $MnTe_2$, $U = 4$ eV was used for calculating total energies of both insulating Mn compounds and metallic bulk Mn because all parameters used for calculating total energy differences need to be the same. If the error in $\mu_{Mn}^{bulk}$, caused by the $U$ parameter, is $\Delta\mu_{\mathrm{Mn}}^{bulk}$, the enthalpies of formation for $MnBi_2Te_4$, $MnBi_4Te_7$, $MnSb_2Te_4$, MnTe, and $MnTe_2$ have a systematic error of $-\Delta\mu_{\mathrm{Mn}}^{bulk}$. However, all these errors do not affect our results for the reasons given below.

For defect formation energy calculations using Eq. 1, only the total chemical potential of Mn, $\mu_{\mathrm{Mn}} + \mu_{\mathrm{Mn}}^{bulk}$, matters and it is not affected by the error in $\mu_{\mathrm{Mn}}^{bulk}$ because the error of $-\Delta\mu_{\mathrm{Mn}}^{bulk}$ in enthalpies of formation of Mn compounds is transferred to $\mu_{\mathrm{Mn}}$ through Eqs. (2), (4), and (6); consequently, the errors in $\mu_{\mathrm{Mn}}$ and $\mu_{\mathrm{Mn}}^{bulk}$ cancel each other.

In fact, the error in $\mu_{\mathrm{Mn}}^{bulk}$ affects only phase diagrams in Figure 2. Specifically, the error in $\mu_{\mathrm{Mn}}^{bulk}$ affects $\mu_{\mathrm{Mn}} = 0$ lines in Figure 2 but does not affect phase boundaries between ternary and binary phases due to the error cancelation. Taking $MnBi_2Te_4$ as an example, the phase boundaries between $MnBi_2Te_4$ and MnTe and $MnTe_2$ are determined by

$$\mu_{\mathrm{Te}} \geq -\frac{2}{3}\mu_{\mathrm{Bi}} + \frac{1}{3}[\Delta H(\mathrm{MnBi_2Te_4}) - \Delta H(\mathrm{MnTe})] \quad (14)$$

and

$$\mu_{\mathrm{Te}} \geq -\mu_{\mathrm{Bi}} + \frac{1}{2}[\Delta H(\mathrm{MnBi_2Te_4}) - \Delta H(\mathrm{MnTe_2})], \quad (15)$$

respectively, following Eqs 2-3. The error is canceled when taking the energy difference between two enthalpies of formation.

Next, we estimate how the error in $\mu_{Mn}^{bulk}$ affects $\mu_{\mathrm{Mn}}$ = 0 lines in phase diagrams in Figure 2. Using $MnBi_2Te_4$ as an example again, the $\mu_{\mathrm{Mn}}$ = 0 line in Figure 2(a) is determined by

$$\mu_{\mathrm{Te}} = -\frac{1}{2}\mu_{\mathrm{Bi}} + \frac{1}{4}\Delta H(\mathrm{MnBi_2Te_4}) \tag{16}$$

(following Eq. 2). Therefore, the error of $-\Delta\mu_{\mathrm{Mn}}^{bulk}$ in $\Delta H(\mathrm{MnBi_2Te_4})$ is scaled down by a factor of 4 for the $\mu_{\mathrm{Mn}}$ = 0 line. We can estimate the error in $\mu_{Mn}^{bulk}$. The calculated $\Delta H(\mathrm{MnTe})$ and $\Delta H(\mathrm{MnTe_2})$ are -1.38 eV and -1.45 eV, respectively. They are lower than their respective measured values of -1.10 eV and -1.30 eV[49] by 0.28 eV and 0.15 eV. Most of the error should come from the artificially high $\mu_{Mn}^{bulk}$. For comparison, the calculated $\Delta H(\mathrm{Bi_2Te_3})$ is -0.97 eV, in good agreement with the experimentally measured value of -1.03 ± 0.1 eV,[50] suggesting that the calculated $\mu_{Te}^{bulk}$ is reasonably accurate. If we assign all the errors in calculated $\Delta H(\mathrm{MnTe})$ and $\Delta H(\mathrm{MnTe_2})$ to $\mu_{Mn}^{bulk}$, we obtain an error < 0.28 eV for $\mu_{Mn}^{bulk}$ and $\Delta H(\mathrm{MnBi_2Te_4})$. Thus, the error in $\mu_{Mn}^{bulk}$ would shift the $\mu_{\mathrm{Mn}}$ = 0 line up by < 0.07 eV in Fig. 2(a). The corrected $\mu_{\mathrm{Mn}}$ = 0 line does not intersect the line segment between points A and B. The same analysis can be applied to $MnBi_4Te_7$ and $MnSb_2Te_4$ and show that the error in $\mu_{Mn}^{bulk}$ does not affect the chemical potential range (the line segment between points A and B in Fig. 2) used in formation energy calculations. This makes sense because the ternary phases likely have phase boundaries with binary phases but not with elemental phases. Therefore, the error in calculating the total energy of bulk Mn does not affect the results of defect/dopant formation energies in $MnBi_2Te_4$, $MnBi_4Te_7$ and $MnSb_2Te_4$.

**ACKNOWLEDGMENTS**

The authors are grateful for the stimulating discussion with Yaohua Liu, Matthew Brahlek, and Hu Miao. This work was supported by the U. S. Department of Energy, Office of Science, Basic Energy Sciences, Materials Sciences and Engineering Division.

Ceramics **47**, 234 (2008).